
\input harvmac

\def\L{\Lambda}

\lref\dv{ R.~Dijkgraaf and C.~Vafa,
Nucl.\ Phys.\ B {\bf 644}, 3 (2002), hep-th/0206255;
Nucl.\ Phys.\ B {\bf 644}, 21 (2002), hep-th/0207106;
hep-th/0208048;
hep-th/0302011.
R.~Dijkgraaf, M.~T.~Grisaru, C.~S.~Lam, C.~Vafa
and D.~Zanon,
hep-th/0211017.
}
\lref\konishione{ K.~Konishi,
Phys.\ Lett.\ B {\bf 135}, 439 (1984).
K.~i.~Konishi and K.~i.~Shizuya,
Nuovo Cim.\ A {\bf 90}, 111 (1985).
}
\lref\CDSW{ F.~Cachazo, M.~R.~Douglas, N.~Seiberg and E.~Witten,
JHEP {\bf 0212}, 071 (2002), hep-th/0211170.
F.~Cachazo, N.~Seiberg and E.~Witten,
JHEP {\bf 0302}, 042 (2003), hep-th/0301006;
JHEP {\bf 0304}, 018 (2003), hep-th/0303207.
}
\lref\MatoneWH{
M.~Matone,
Nucl.\ Phys.\ B {\bf 656}, 78 (2003), hep-th/0212253.
}
\lref\WittenYE{ E.~Witten,
hep-th/0302194.
}
\lref\EtingofDD{
P.~Etingof and V.~Kac,
math.qa/0305175.
}
\lref\MatoneNA{
M.~Matone and L.~Mazzucato,
JHEP {\bf 0307}, 015 (2003), hep-th/0305225.
}
\lref\KrausJF{
P.~Kraus and M.~Shigemori,
JHEP {\bf 0304}, 052 (2003), hep-th/0303104.
L.~F.~Alday and M.~Cirafici,
JHEP {\bf 0305}, 041 (2003), hep-th/0304119.
P.~Kraus, A.~V.~Ryzhov and M.~Shigemori,
JHEP {\bf 0305}, 059 (2003), hep-th/0304138.
M.~Aganagic, K.~Intriligator, C.~Vafa and N.~P.~Warner,
hep-th/0304271.
F.~Cachazo,
hep-th/0307063.
}
\lref\MatoneRX{
M.~Matone,
Phys.\ Lett.\ B {\bf 357}, 342 (1995), hep-th/9506102;
JHEP {\bf 0104}, 041 (2001), hep-th/0103246;
Phys.\ Lett.\  {\bf 514}, 161 (2001), hep-th/0105041.
}
\lref\DoreyZQ{
N.~Dorey, T.~J.~Hollowood and V.~V.~Khoze,
JHEP {\bf 0103}, 040 (2001), hep-th/0011247.
F.~Fucito, J.~F.~Morales and A.~Tanzini,
JHEP {\bf 0107}, 012 (2001), hep-th/0106061.
 T.~J.~Hollowood,
JHEP {\bf 0203}, 038 (2002), hep-th/0201075.
}
\lref\HollowoodNG{ T.~J.~Hollowood, V.~V.~Khoze and G.~Travaglini,
JHEP {\bf 0105}, 051 (2001), hep-th/0102045.
}
\lref\HollowoodZV{ T.~J.~Hollowood,
Nucl.\ Phys.\ B {\bf 639}, 66 (2002), hep-th/0202197.
}
\lref\NekrasovQD{ N.~A.~Nekrasov,
hep-th/0206161.
}
\lref\NakajimaPG{
H.~Nakajima and K.~Yoshioka,
math.ag/0306198.
N.~Nekrasov and A.~Okounkov,
hep-th/0306238.
}
\lref\HollowoodQN{ T.~J.~Hollowood, V.~V.~Khoze, W.~J.~Lee and
M.~P.~Mattis,
Nucl.\ Phys.\ B {\bf 570}, 241 (2000), hep-th/9904116.
}
\lref\KonishiTS{
K.~Konishi and A.~Ricco,
hep-th/0306128.
}
\lref\LosevTP{
A.~Losev, N.~Nekrasov and S.~L.~Shatashvili,
Nucl.\ Phys.\ B {\bf 534}, 549 (1998), hep-th/9711108;
hep-th/9801061.
}
\lref\MooreDJ{
G.~W.~Moore, N.~Nekrasov and S.~Shatashvili,
Commun.\ Math.\ Phys.\  {\bf 209}, 97 (2000), hep-th/9712241;
Commun.\ Math.\ Phys.\ {\bf 209}, 77 (2000), hep-th/9803265.
N.~Nekrasov and A.~Schwarz,
Commun.\ Math.\ Phys.\  {\bf 198}, 689 (1998), hep-th/9802068.
}
\lref\FlumeAZ{
R.~Flume and R.~Poghossian,
hep-th/0208176.
U.~Bruzzo, F.~Fucito, J.~F.~Morales and A.~Tanzini,
JHEP {\bf 0305}, 054 (2003), hep-th/0211108.
}

\Title{\vbox{\baselineskip11pt\hbox{hep-th/0307130}
\hbox{DFPD/03/TH/22} \hbox{SISSA 47/2003/EP} }} {\vbox{
\centerline{On the Chiral Ring of ${\cal N}=1$}
\smallskip
\centerline{Supersymmetric Gauge Theories}
 \vskip
1pt }}
\smallskip
\centerline{Marco Matone$^1$ and Luca Mazzucato$^2$}
\smallskip
\bigskip
\centerline{$^1$Dipartimento di Fisica ``G. Galilei'', Istituto
Nazionale di Fisica Nucleare,} \centerline{Universit\`a di Padova,
Via Marzolo, 8 -- 35131 Padova, Italy}
\medskip
\centerline{$^2$International School for Advanced Studies,
Trieste, Italy}
\bigskip
\vskip 1cm
 \noindent
We consider the chiral ring of the pure
${\cal N}=1$ supersymmetric gauge theory with $SU(N)$ gauge group and show that
the classical relation $S_{cl}^{N^2}=0$ is modified to the exact quantum relation
$(S^N-\Lambda^{3N})^N=0$.

\vskip 0.5cm

\Date{July 2003}
%
\baselineskip14pt

%
%


Recently, much attention has been devoted to the study of four
dimensional ${\cal N}=1$ supersymmetric gauge theories, due to the
gauge theory/matrix model correspondence \dv. This result has been
clarified from the field theoretical point of view in
\CDSW\WittenYE\ by considering the chiral ring of the gauge theory
and a generalization of the Konishi anomaly \konishione.
Furthermore, the relation with ${\cal N}=2$ supersymmetric gauge
theories led to an underlying duality in the ${\cal N}=1$ theory
\MatoneWH. This duality is strictly related to scaling properties
of the matrix model free energy. The latter has been also useful
in investigating the exact structure of the free energy, leading
to the appearance of new bilinear terms depending on an odd
integer \MatoneNA. In this respect we note that, in particular
theories, the are interesting questions concerning the
contributions at order $S^h$, with $h$ the dual Coxeter number
\KrausJF.

In this note we will consider the chiral ring structure for pure ${\cal N}=1$
$SU(N)$ gauge theory and we will
argue that the classical relation \CDSW
\eqn\one{ S_{cl}^{N^2}=0, }
where $S=-{1\over 32\pi^2}\Tr\, W_\alpha W^\alpha$ is the glueball superfield,
gets modified to the exact quantum relation
\eqn\proptwo{ \left(S^N -
\L^{3N}\right)^N=0. }

Let us consider the gaugino condensate in the case of $SU(N)$.
Properties of the trace for $SU(N)$ show that classically also the
following relation \CDSW\ \eqn\two{ S_{cl}^N=\{\overline
Q_{\dot\alpha},X^{\dot\alpha}\},} holds, whose generalization to
gauge groups $Sp(N)$ and $SO(N)$ has been derived by Witten in
\WittenYE. Subsequently, in \EtingofDD\ it has been verified the
structure of the classical ring for the exceptional Lie group
$G_2$ conjectured in \WittenYE. Instantons modify Eq.\two\ to the
exact operator relation \CDSW\ \eqn\three{
S^N=\Lambda^{3N}+\{\overline Q_{\dot\alpha},X^{\dot\alpha}\},}
that generalizes to other groups \WittenYE. Similarly, also
$S^{N^2}=0$ receives instanton corrections. In particular,
consistency with the above finding implies that there is the exact
operator relation \eqn\four{ {\cal P}(S^N,\Lambda^{3N})=0, } where
${\cal P}\equiv(S^N-\Lambda^{3N})P(S^N,\Lambda^{3N})$ with $P$ a
homogeneous polynomial of degree $N-1$ with a non--zero
coefficient of $(S^N)^{N-1}$, whose precise form is unknown \CDSW.

Fermi statistics requires
attention in studying the quantum properties of $S$ and its powers
as these need to be defined by point splitting.
Instanton calculations indicate that one may obtain a well defined
field constructed out of $S$. For $SU(N)$ one obtains (see
\KonishiTS\ for a recent discussion) \eqn\by{ \langle S^N
\rangle=\Lambda^{3N}. } This result needs to be specified. There
are two ways to calculate the gluino condensate. One is based on
the weak--coupling instanton (WCI) calculations, giving the above
result, whereas with the strong--coupling instanton (SCI)
calculations the right hand side of \by\ is replaced by
$2[(N-1)!(3N-1)]^{-1/N}\L^{3N}$. However, it turns out that
cluster decomposition does not hold in the SCI \HollowoodQN.
Furthermore, it has been observed that on ${\bf R}^3\times S^1$
the results do not depend on the radius of $S^1$,
so that one ends up with ${\bf R}^4$ in the infinite radius limit,
or more precisely ${\bf R}^3\times \hat {\bf R}$, where
$\hat{\bf R}\doteq{\bf R}\cup\{\infty\}$ is the one point compactification
of ${\bf R}$.
It is interesting to note that, in the case of ${\cal
N}=2$, it has been shown that the instanton moduli space admits a
compactification induced by the noncommutative theory. This property is
intrinsic to ${\cal N}=2$, and so, even if the
field content in the strong coupling region has not emerged yet, it should have a counterpart
in a field theoretic derivation of the
expansion of the dual SW prepotential ${\cal F}_D$ near the critical points $u=\pm\Lambda^2_{SW}$.
On the other hand, this region is the one where breaking the ${\cal N}=2$ theory one obtains
the ${\cal N}=1$ result \by.
This would suggest that the equivalence between the calculation of the gluino condensate
in ${\bf R}^4$, made by the WCI calculations, and the evaluation of the
gluino condensate in ${\bf R}^3\times S^1$,
may be connected with the compactification, induced by noncommutative
geometry, of the instanton moduli spaces in ${\cal N}=2$.
In this respect, it is worth recalling that in the infrared regime
low--energy dynamics of noncommutative ${\cal N}=2$ supersymmetric
$U(N)$ Yang--Mills theories, the $U(1)$ decouples and the $SU(N)$ is described
by the commutative SW theory \HollowoodNG.

Using the recursion relations for the instanton contributions, a
Deligne--Knudsen--Mumford (DKM) like compactification of instanton
moduli space was derived in \MatoneRX. A remarkable property of
the DKM stable compactification is a sort of regularization as
punctures never collide in the degeneration limit. This is at the
basis of the recursive structure. It also turns out that, in a
different approach, the existence of a nilpotent fermionic
symmetry implies a BRST operator that leads to localization of
integrals on the instanton moduli spaces \DoreyZQ. This is related
to Nekrasov's project, formulated in a series of papers
\LosevTP\MooreDJ. He and collaborators considered localization
onto the instanton moduli space, and introduced both the
$Q$--operator, which makes use of spacetime rotations in addition
to the global gauge transformations, as well as the concept of the
noncommutative instantons. Remarkably, this culminated in
\NekrasovQD, with the explicit evaluation, for any classical gauge
group, of the instanton moduli space integrals (see also
\NakajimaPG\FlumeAZ).

The above discussion naturally leads to consider nonperturbative configurations
of ${\cal N}=1$ supersymmetric gauge theory on ${\bf R}^3\times \hat{\bf R}$ rather
than ${\bf R}^4$. As argued by Witten \WittenYE, this essentially avoids possible
infrared divergences due to the fact that the calculations are performed by choosing
a perturbative vacuum
which is different from the true one. This one--point compactification can be seen
as a way to impose boundary conditions on nonperturbative configurations rather than a change
of topology of the space where the gauge theory lives.
This is of interest for the definition of the gluino condensate.

We also note that the point splitting is a quantum operation which leads to a modification of the
classical operator definition. Even if this operation is sensible
to UV physics, the above remarks indicate that topological
properties of the space--time and the possible connection to
noncommutative geometry may lead to some UV/IR mixing related to
the underlying fermionic nature of the gaugino condensate. Moreover,
there are some analogies between the matrix
model formulation and the noncommutative theory.

Let us go back to the analysis of the chiral ring. The above
discussion shows that we can use instanton results in order to
define \eqn\split{{\cal O}_\Lambda \doteq S^N-\lim_{x_i\to
x_j\atop \forall ij}\langle S(x_1)\ldots S(x_N)\rangle
=S^N-\Lambda^{3N}.} Note that the $X^{\dot\alpha}$ in \two\ and
\three\ can differ only by a chiral operator: dimensional analysis
and $R$--symmetry forbid the existence of terms $\{\overline
Q_{\dot\alpha},\delta X^{\dot\alpha}\}$ that vanish as
$\Lambda\to0$. The correction from $S_{cl}^N$ to $S^N$ concerns a
redefinition of the glueball superfield and not $\{\overline
Q_{\dot\alpha},X^{\dot\alpha}\}$, that is \eqn\cucchi{
S_{cl}^N=S^N-\Lambda^{3N}. } Therefore, the basic observation is
that it is the $N$--th power of the glueball superfield that gets
quantum corrections. For these reasons we used the notation ${\cal
O}_\Lambda$ in \split\ instead of $S_{cl}^N$. However, since
$S_{cl}^{N^2}=0$ was derived as an identity, and since, as we
said, $\{\overline Q_{\dot\alpha},X^{\dot\alpha}\}$ does not
receive quantum corrections, it follows by \one\ and \two\
\eqn\po{ \{\overline Q_{\dot\alpha},X^{\dot\alpha}\}^N=0. } On the
other hand, being \eqn\peppe{ {\cal O}_\Lambda=\{\overline
Q_{\dot\alpha},X^{\dot\alpha}\}, } we have \eqn\funziona{ {\cal
O}_{\Lambda}^N=0, } that is Eq.\proptwo, as promised.

We conclude this note by observing that the emerging structure is
reminiscent of the property of forms in a $(N-1)$--dimensional space.
To realize the similarity let us write
\eqn\unoforma{
\omega\doteq\{\overline Q_{\dot\alpha},X^{\dot\alpha}\},
}
where $\omega$ is a one--form on a $(N-1)$--dimensional
space. Then
\eqn\sopravolume{
\{\overline Q_{\dot\alpha},X^{\dot\alpha}\}^N=\wedge_{k=1}^N\omega=0,
}
leading to a structure which should be further investigated.

\vskip 20pt

\noindent {\bf Acknowledgements}. It is pleasure to thank L. Alday, M. Cirafici, M. Tsulaia
and G. Travaglini for discussions. Work partially supported by the
European Community's Human Potential Programme under contract
HPRN-CT-2000-00131 Quantum Spacetime.

\listrefs

\end